\documentstyle[prl,aps,preprint,tighten,epsf]{revtex}

\begin{document}
\draft
\title{Metal-insulator transition in highly conducting oriented polymers }
\author{N. Dupuis }
\address{Laboratoire de Physique des Solides,
Universit\' e Paris-Sud,
91405 Orsay, France   }
\maketitle
\begin{abstract}
We suggest that highly conducting oriented polymers with a fibril structure
can be modeled by a regular lattice of disordered metallic wires with a random
first-neighbor interwire coupling which mimics the cross-links between
fibrils. We show that such a model can be described by a
non-linear sigma model. Within a one-loop
self-consistent approximation, we determine the position of the
metal-insulator transition as a function of interwire cross-links
concentration, interwire coupling and number of polymer chains in a wire. 
\end{abstract}
\pacs{PACS numbers: 71.20Hk, 71.30.+h, 72.15.Rn}

Highly conducting doped polymers based on polyacetylene, polypyrrole, and
polyaniline have recently attracted considerable interest. Due to the advent
of improved chemical processing, their room-temperature can be comparable to
that of copper. A $T$-independent Pauli susceptibility, a linear $T$-dependence
of thermoelectric power and a large negative microwave dielectric constant
further suggest that highly conducting polymers are intrinsically
metallic. However, in contrast to traditional metals, their conductivity
decreases with temperature. This latter property is usually explained by the
fact that highly conducting polymers are close to a metal-insulator transition
(MIT) driven by disorder \cite{poly}. The decrease of the 
conductivity with $T$ then
results from phonon-controlled localization effects. 
Motivated by these experimental
results, many explanations of the transport properties of highly conducting
polymers are based on dirty quasi-one-dimensional conductors (i.e. weakly
coupled chain systems) \cite{Q1D}. The assumption of a periodic 
arrangement of the 1D
polymers chains which underlies such explanations is however quite
unrealistic. Prigodin and Efetov (PE) have recently proposed a model which
takes into account the irregular structure of the polymers
\cite{Prigodin93}. In highly conducting polymers, single chains are coupled 
into fibrils which are bent in space in a very complicated way. PE model
the fibrils by weakly disordered metallic wires. The cross-links between the
fibrils are described by interwire junctions. In the absence of junctions,
all electronic states of 
the wires are localized by any weak disorder. PE have shown that the
interwire junctions lead to a MIT and determined the 
position of the transition and the critical behavior. 

We suggest in this Letter that highly conducting oriented polymers with a
fibril structure can be modeled by a regular lattice of disordered metallic
wires with a random first-neighbor interwire coupling. The
case where the wires contain a single chain has been recently studied
numerically by Zambetaki, Economou and Evangelou (ZEE) \cite{Zambetaki}. 
We show that the MIT in this regular lattice of
randomly coupled wires can be described by a non-linear sigma model 
(NL$\sigma $M) which we solve in a one-loop self-consistent
approximation. For a large number $M$ of 1D chains in the fibrils, we recover
the main result of PE: an increase of the interwire coupling $J$
favors a delocalization of the electronic states. However we obtain a
different critical behavior and the position of the transition
differs from the result of PE. When $M=1$, we show that a small
value of $J$ favors the metallic state while a large value of $J$ induces
localization in agreement with the numerical calculation of ZEE.

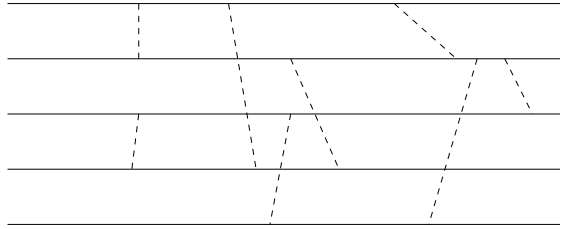
\begin{figure}
\caption{(a) Schematic representation of a random network of metallic
disordered wires. Each intersection between solid lines represents an
interwire junction. (b) Stretching the wires, we obtain a regular
lattice. The interwire junctions become interwire couplings (dashed lines). }
\label{Fig1}
\end{figure}

We consider PE's model in the case of strongly oriented
polymers and make the following  Gedanken experiment: we
stretch the wires in such a way that they form a regular (square) lattice of
straight and parallel wires (Fig.\ \ref{Fig1}). 
The interwire junctions now correspond to couplings between neighboring 
wires. In the case of a strongly oriented polymer, these
couplings will have a short range. It is then possible, without inducing any
qualitative change, to consider a simpler
model where the couplings are allowed only between first-neighbor wires and
in a direction which is perpendicular to the axes of the wires. To further 
specify the model, we assume that at each position $x$ of a given wire, the
coupling strength with a neighboring wire is equal to $J$ with probability
$c$ (presence of an interwire junction) and vanishes with probability $1-c$
(absence of a junction). For $M=1$, this model exactly corresponds to the
one studied numerically by ZEE. It is clear that the model is meaningful
only if the probability to have a junction at a given position along a wire
is weak ($c\ll 1$). As pointed out in Ref.\ \cite{Prigodin93},
the existence of a delocalized phase in this model is a highly non-trivial
phenomenon. Indeed, the random interwire coupling increases the dimensionality
(which favors the delocalized phase) but is also an additional source of
scattering. 

We first consider the case where each fibril reduces to a single chain
($M=1$). We will show below how the results can be easily extended to an
arbitrary number of channels. We note $\tau $ the elastic scattering time
which results from intrachain disorder.  In the absence
of interchain coupling the electronic states in the chain are localized with
a localization length $R_0\sim l$ where $l=v_F\tau $ is the mean free path
and $v_F$ the Fermi velocity along the chain. 
The mean-value of the coupling between
neighboring chains is $\bar t_\perp 
=cJ$. Deviations from this mean-value are taken into account via a random
hopping $V^\perp _{l,l'}(x)$ with second cumulant equal to
\begin{eqnarray}
\langle V^\perp _{l_1,l_2}(x) V^\perp _{l_1',l_2'}(x') \rangle
&=& \tilde t_\perp \delta (x-x') \bar \delta _{l_1,l_2\pm 1} 
\nonumber \\ && \times (\delta _{l_1,l_1'}
\delta _{l_2,l_2'}+ \delta _{l_1,l_2'} \delta _{l_2,l_1'}) \,,
\label{cumu}
\end{eqnarray}
where $\tilde t_\perp =ac(1-c)J^2$. 
$a$ is the lattice spacing along the chains axis. $l\equiv (l_y,l_z)$
($l_y,l_z$ integer) determines the position of a given chain. $\bar \delta
_{l_1,l_2\pm 1}$ means that $l_1$ and $l_2$ are first-neighbors. Higher
order cumulants of the random variable $V^\perp $ are neglected. 
Averaging over the intrachain disorder and
the random coupling $V^\perp $ is performed by introducing
$N$ replica of the system. The partition function can be written as a
functional integral over Grassmann variables ${\psi ^\alpha }^{(*)}$  
($\alpha =1,...,N$). The action $S$ contains two terms: i) $S_0$
describing a weakly coupled chains system with a dispersion law
$\epsilon_{\bf k}=v_F(\vert k_x\vert -k_F)-2\bar t_\perp (\cos (k_yd)+\cos
(k_zd))$ in  presence of intrachain disorder (with elastic scattering time
$\tau $)  ($k_F$ is the
Fermi momentum of a single chain, and $d$ the spacing between chains); ii)
$S_{\rm dis}^\perp $ coming from the  random coupling $V^\perp $. Using
(\ref{cumu}), we obtain  
\begin{eqnarray}
S_{\rm dis}^\perp  
&=& -{\tilde t_\perp  \over 2} \sum _{\alpha ,\beta ,l,l'} 
\bar \delta _{l,l'\pm 1} 
\int dx \,d\tau \,d\tau ' \nonumber \\ && \times 
\Bigl \lbrack {\psi ^\alpha _l}^*(x,\tau ) 
{\psi ^\beta _l}^*(x,\tau ') \psi ^\beta _{l'}(x,\tau ') 
 \psi ^\alpha _{l'}(x,\tau ) \nonumber \\ 
&& -  {\psi ^\alpha _l}^*(x,\tau ) 
\psi ^\beta _l(x,\tau ') {\psi ^\beta _{l'}}^*(x,\tau ') 
 \psi ^\alpha _{l'}(x,\tau ) \Bigr \rbrack \,,
\label{Sperp}
\end{eqnarray}
where $\tau ,\tau '\in \lbrack 0,1/T\rbrack $ are imaginary times. 
Eq.\ (\ref{Sperp}) shows that the averaging over $V^\perp $
generates hoppings of particle-particle or particle-hole pairs between
neighboring chains. 

In the following, we decouple the quartic term $S_{\rm dis}^\perp $ by
means of a Hubbard-Stratonovich transformation and obtain the effective
action of the low-energy fluctuations of the auxiliary field around its
saddle point value. The parameters of this effective action are
expressed as a function of $\tilde t_\perp $ and the correlation functions
obtained from an action $S_0''$ which is related to $S_0$ in a simple
way (see Eqs.\ (\ref{SQ1},\ref{Rcal}) below) \cite{Boies95}. 

In order to take into
account on the same footing the particle-particle and particle-hole
channels, we introduce spinors $\phi ,\bar \phi $ defined by 
\begin{equation} 
\phi _{ln}^\alpha (x)={1 \over \sqrt{2}} 
\left (
\begin{array}{l}
 {\psi ^\alpha _{ln}}^*(x) \\ \psi ^\alpha _{ln}(x)
\end{array} \right )
\end{equation}
and $\bar \phi =(C\phi )^T$ where $C$ is the charge conjugaison operator
\cite{Belitz94}.
$\psi ^\alpha _{ln}(x) \equiv \psi ^\alpha _l(x,\omega _n)$ are the
Fourier transformed fields with respect to the imaginary time and $\omega
_n=\pi T(2n+1)$. The action $S_{\rm dis}^\perp $ is rewritten as
\begin{equation}
S_{\rm dis}^\perp \lbrack \bar \phi ,\phi \rbrack =\int dx\,
{\rm Tr} \lbrack B(x) \tilde t_\perp  
B(x) \rbrack  \,,
\label{Sperp1}
\end{equation}
where we have introduced the matrix field $B^{\alpha \beta }_{lnm}(x)=\phi
^\alpha _{ln}(x) \otimes \bar \phi ^\beta _{lm}(x)$. 
In (\ref{Sperp1}), $\tilde t_\perp $ should be
understood as the matrix $\tilde t_{\perp l,l'}=\tilde t_\perp \bar \delta 
_{l,l'\pm 1}$ (diagonal in the indices $\alpha, n,i$ where $i=1,2$ refers to
the two components of the spinors). 
Tr denotes the trace over all discrete indices. 
Introducing an auxiliary matrix field $Q^{\alpha \beta
}_{lnm}(x)$ to decouple the quartic term (\ref{Sperp1}), we write the
partition function as
\begin{eqnarray}
Z &=& \int {\cal D}\bar \phi {\cal D}\phi \, 
e^{-S_0\lbrack \bar \phi ,\phi \rbrack } \nonumber \\ && \times 
\int {\cal D}Q \, e^{- \int dx\,  \bigl \lbrack 
{\rm Tr}  \lbrack Q(x)
\tilde t^{-1}_\perp Q(x) \rbrack 
+2i{\rm Tr} \lbrack B(x)Q(x) \rbrack \bigr \rbrack  } \,, 
\end{eqnarray}
where $\tilde t^{-1}_\perp $ is the inverse matrix of $\tilde t_\perp $. (One
can verify that different decouplings of $S_{\rm dis}^\perp $ would not lead
to any non-trivial phenomena.) The
auxiliary field has the same structure as the field $B(x)$ and therefore
satisfies the conditions $Q^+=C^TQ^TC=Q$ \cite{Belitz94}.

We first determine the value of $Q$ in the saddle point
approximation. Assuming a solution of the form $_{ij}(Q^{\rm SP})^{\alpha
\beta } 
_{lnm}(x)=Q_0 \delta _{\alpha ,\beta }\delta _{n,m}\delta _{i,j}$ we obtain the
saddle point equations
\begin{eqnarray}
Q_0 &=& {i \over 2} \tilde t_\perp ({\bf q}_\perp =0) {1 \over {LN^2_\perp }}
\sum _{\bf k} G^{\rm SP}_n({\bf k}) \,, \nonumber \\
G^{\rm SP}_n({\bf k}) &=& \Bigl \lbrack i\omega _n -\epsilon _{\bf k} +{i 
\over {2\tau }}{\rm sgn}(\omega _n)+2iQ_0 \Bigr \rbrack ^{-1} \,, 
\label{SPeq}
\end{eqnarray}
where $\tilde t_\perp ({\bf q}_\perp )=2\tilde t_\perp (\cos (q_yd)+\cos
(q_zd))$ is the Fourier transform of $\tilde
t_\perp \bar \delta _{l,l'\pm 1}$. $N_\perp ^2$
is the total number of chains and $L$ the length of the chains. 
Eq.\ (\ref{SPeq}) is 
obtained using the Born approximation for the single-particle Green's
function corresponding to $S_0$. From (\ref{SPeq}), we obtain
$Q_0=(1/4\tau '){\rm sgn}(\omega _n)$ with $1/\tau '=8\pi N_1(0)\tilde
t_\perp $ where $N_1(0)=1/\pi v_F$ is the density of states of a single
chain. Within the saddle point approximation, we therefore obtain a change
of the elastic scattering time due to the random interchain hopping. The total
scattering rate becomes $1/\tau ''=1/\tau +1/\tau '$. 

We now consider the fluctuations around the saddle point solution. As in
the standard localization problem \cite{Belitz94}, the
low-energy fluctuations correspond to spatially slowly varying fields which
satisfy the constraints $Q^2={Q^{\rm SP}}^2$ and ${\rm Tr}\,Q=0$. 
Following Ref.\
\cite{Belitz94}, we shift the field according to $Q\to Q+Q^{\rm SP}-\Omega /2$,
with $_{ij}\Omega ^{\alpha \beta }_{lnm}(x)=\delta _{\alpha ,\beta }\delta
_{n,m}\delta _{i,j}\omega _n$, and expand the action to lowest order in
$\Omega $ and $Q$. Using the saddle point conditions (\ref{SPeq}) we obtain
the effective action of the $Q$ field 
\begin{equation}
S\lbrack Q \rbrack = \int dx \, {\rm Tr} \lbrack Q(x) \tilde t^{-1}_\perp
Q(x) -\Omega \tilde t^{-1}_\perp Q(x) \rbrack 
-\int dx_1  dx_2 \, _{ji}Q^{\beta \alpha }_{l_1mn} (x_1) 
_{ij}{\cal R}^{\alpha \beta } _{nm} (x_1l_1,x_2l_2) 
_{ij}Q^{\alpha \beta }_{l_2nm} (x_2) \,, 
\label{SQ1}
\end{equation}
where a sum over repeated indices is implied. ${\cal R}^{\alpha \beta }
_{nm}$ is defined by 
\begin{eqnarray}
&& {\cal R}^{\alpha \beta } _{nm} (1,2)= 
\left (
\begin{array}{lr}
R^{\alpha \beta }_{nm}(2,1 \vert 1,2) & 
R^{\alpha \beta }_{nm}(2,2 \vert 1,1) \\
R^{\alpha \beta }_{nm}(1,1 \vert 2,2) & 
R^{\alpha \beta }_{nm}(1,2 \vert 2,1) \\
\end{array}
\right ) \,,  \label{Rcal} \nonumber \\  &&
R^{\alpha \beta }_{nm}(x_1l_1,x_2l_2\vert x_3l_3,x_4l_4)= 
\langle \psi ^\alpha _{l_1n}(x_1) \psi ^\beta _{l_2m}(x_2)
{\psi ^\beta }^*_{l_4m}(x_4) {\psi ^\alpha }^*_{l_3n}(x_3) 
\rangle _{\tilde S_0} \,. 
\end{eqnarray}
The action  $\tilde S_0$ is equal to $S_0+2i\int dx\, {\rm Tr}\lbrack
B(x)(Q^{\rm SP}-\Omega /2)\rbrack $.
For a system with time reversal symmetry, the Fourier transform ${\cal
R}^{\alpha \beta }_{nm}({\bf q})$ has all its components equal to $R^{\alpha
\beta }_{nm}({\bf q})$. In the diffusive regime, it is easy to obtain the
expression of $R^{\alpha \beta }_{nm}({\bf q})$ from the action $\tilde
S_0$. In the replica limit ($N\to 0$), we have for the diffusive (Goldstone)
modes ($\omega _n\omega _m<0$)
\begin{equation}
R^{\alpha \alpha }_{nm}({\bf q})^{-1} \Bigr \vert _{\rm diff}={1 \over
{2\pi N_1(0)}} 
\biggl ( {1 \over {\tau '}} +v_F^2\tau ''q_x^2  
+8\bar t^2_\perp \tau '' 
\Bigl (\sin ^2(q_yd/2)+\sin ^2(q_zd/2) \Bigr ) \biggr ) \,.
\label{Rdiff}
\end{equation}
The saddle point $Q_0$ introduces a ``mass'' term in the diffusive
propagator. The exact propagator $R^{\alpha \alpha }_{nm}({\bf q})$ 
can be expressed as in
(\ref{Rdiff}) but with renormalized diffusive coefficients $\tilde D''$ and
$\tilde D''_\perp $ replacing the bare coefficients $v_F^2\tau ''$ and $8\bar
t^2_\perp \tau ''$, respectively. Notice that the ``mass'' term $1/\tau '$ 
plays the same role as a finite frequency. $R^{\alpha \alpha }_{nm}({\bf
q})$ can be obtained either from $\tilde S_0$ or from $S_0''$ but at a finite
frequency $1/\tau '$. Here $S_0''$ is the action obtained from $S_0$ by the
replacement $\tau \to \tau ''$. Eqs.\ (\ref{SQ1},\ref{Rcal})  yield
(after a rescaling of the fields in order to have $Q^2=\underline 1$ with 
$\underline 1$ the unit matrix) 
\begin{eqnarray}
S\lbrack Q\rbrack &=& {\pi \over 8}N_1(0) \sum _{\bf q} \biggl  ( \tilde D''
q_x^2 + \biggl (\tilde D''_\perp +{1 \over {\tau '}} \biggr )
\Bigl (\sin ^2(q_yd/2) 
+\sin ^2(q_zd/2)\Bigr  ) \biggr ) {\rm Tr} \lbrack Q({\bf q})
Q(-{\bf q}) \rbrack \nonumber \\ &&
 -{\pi \over 2} N_1(0) {\rm Tr}\lbrack \Omega Q({\bf q}=0) \rbrack \,.
\label{SQ2}
\end{eqnarray}
Notice that the ``mass'' term $1/\tau '$ in the propagator
$R^{\alpha \alpha }_{nm}({\bf q})$ is canceled by the uniform part of
${\rm Tr}\lbrack Q(x)\tilde t^{-1}_\perp Q(x)\rbrack $ (see Eq.\
(\ref{SQ1})), i.e. the low-energy 
excitations correspond to diffusion (Goldstone) modes as expected from
general symmetry arguments \cite{Belitz94}. The effective
transverse diffusion coefficient appearing in (\ref{SQ2}) is the sum of two
terms. $\tilde D_\perp '' \propto \bar t^2$ is generated by single 
particle interchain hopping ($\bar t_\perp $) and $1/\tau '\propto \tilde
t_\perp $ comes from particle-particle or
particle-hole pair-hopping as described by (\ref{Sperp}). 
The effective action (\ref{SQ2}), together with the
constraints on the field $Q$, corresponds to a NL$\sigma $M for a system of
weakly coupled chains. Expanding the $\sin ^2$ in (\ref{SQ2}) to lowest
order (which amounts to keeping only the lowest order gradient terms) would
yield the standard effective action of an anisotropic disordered system
\cite{Wolfle84}.  Although this would be sufficient to obtain the critical
behavior at the MIT, it is necessary to use the action (\ref{SQ2}) to
determine the correct position of the transition. Anderson localization in
weakly coupled chains systems has been studied by Prigodin and Firsov (PF)
\cite{Prigodin83} using a self-consistent diagrammatic treatment introduced
by Vollhard and W\"olfle \cite{Vollhard80}. This latter approach is
equivalent to a one-loop self consistent approximation of the effective
action (\ref{SQ2}). For a system of weakly coupled chains with interchain
coupling 
$t_{\perp {\rm eff}}$ and intrachain elastic scattering time $\tau _{\rm
eff}$, PF found that there is a MIT when $2t_{\perp {\rm eff}}\tau _{\rm eff}
=\gamma $ with $\gamma $ a constant of order unity. The relation $t_{\perp
{\rm eff}}\tau _{\rm eff}={\rm const}$ has been confirmed numerically
\cite{Dorokhov83}. Since our final results will strongly depend on the
precise value of $\gamma $ when $M=1$, we will consider $\gamma $ as an
adjustable parameter in the following. 
In the case of weakly coupled wires, the position of the
MIT is determined by $2Mt_{\perp {\rm eff}}\tau _{\rm eff} =\gamma $ with
$M$ the number of channels in the wires. This dependence on $M$ can be 
obtained from standard scaling arguments \cite{Dupuis92}. Applying
PF's results to the action (\ref{SQ2}), we obtain that the position of the
MIT is determined by
\begin{equation}
\left ( \tilde D''_\perp +{1 \over {\tau '}} \right ) 
{{\tilde D''} \over {v_F^2}} ={{2\gamma ^2} \over {M^2}} \,.
\label{MIT}
\end{equation}

The last step is to obtain the diffusion coefficients $\tilde D''$ and
$\tilde D_\perp ''$. As mentioned above, these coefficients can be obtained
from $S_0''$ but at a finite frequency $1/\tau '$. Again, we have to
calculate diffusive coefficients in a weakly coupled chains (or wires)
system so that we can use PF's results. The position of the MIT is
determined by $2\bar t_\perp \tau ''=\gamma $ ($2M\bar t_\perp \tau ''=\gamma
$ in the case of wires). In the localized phase ($2\bar t_\perp \tau
''<\gamma $), the localization lengths are $\xi =R_0''/\vert \epsilon \vert $
and  $\xi _\perp =\xi l_\perp ''/l''$ with $l''=v_F\tau ''$. $l_\perp
''=\sqrt{2}\bar t_\perp \tau '' d$ and $\xi _\perp $ are the mean free path
and localization length, respectively, in the transverse directions. 
In the case of wires, one should replace the
one-dimensional localization length $R_0''\sim l''$ by the localization length
of an isolated wire $R_0''\sim Ml''$. $\vert \epsilon \vert = 1-(2M/\gamma )
\bar t_\perp \tau ''$ in the localized phase and vanishes in the metallic
phase. The finite frequency diffusion coefficients are given by
\cite{Vollhard80} 
\begin{equation}
D''(\omega _\nu ) = {{v_F^2\tau ''} \over {1+{{{l''}^2} \over {\xi ^2\vert
\omega _\nu \vert \tau ''}}}} \,;\,\,\, 
D''_\perp (\omega _\nu ) = {{8\bar t_\perp ^2\tau ''} \over {1+{{{l''}^2} \over
{\xi ^2\vert \omega _\nu \vert \tau ''}}}} \,. 
\label{CD}
\end{equation}
$\tilde D''$ and $\tilde
D_\perp ''$ are obtained from (\ref{CD}) with $\vert \omega _\nu \vert
=1/\tau '$. From (\ref{MIT},\ref{CD}), we obtain the position of the MIT in the
system of randomly coupled wires:  
\begin{equation}
{{\tau ''/\tau '} \over {1+{{\vert \epsilon \vert ^2 \tau '} \over {M^2\tau
''}}}} +{{8\bar t_\perp ^2\tau ''^2} \over {\left ( 1+{{\vert \epsilon \vert
^2 \tau '} \over {M^2\tau ''}} \right )^2 }} ={{2\gamma ^2} \over {M^2}} \,.
\label{MIT1}
\end{equation}
Following Refs.\ \cite{Prigodin93,Zambetaki}, we introduce the number of
junctions per unit of length, $p=4c/a$, the number of junctions within the
wire localization length, $\rho =pR_0$, and the dimensionless interwire
coupling $\alpha =(\pi aN_1(0)J)^2$. For $c\ll 1$, (\ref{MIT1}) can be
rewritten as
\begin{eqnarray}
&& {1 \over {16\alpha }}+{A \over 2} -{{\gamma ^2} \over {M^2}}A^2 =0 \,, 
\nonumber  \\ &&
A=1+{M \over {2\rho \alpha }} +{1 \over {M^2}} \biggl (1+
{M \over {2\rho \alpha }} -{M \over {4\gamma \sqrt{\alpha }}} \biggr )^2 \,.
\end{eqnarray}

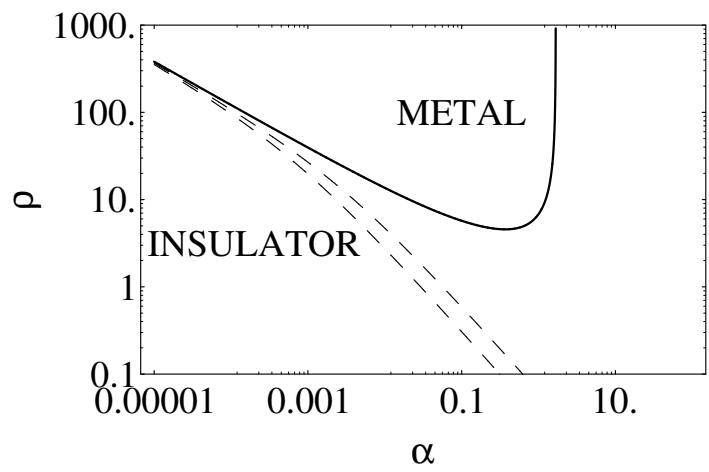
\begin{figure}
\caption{Phase diagram in the region
$c\ll 1$ for $M=1$ (solid line), $M=10$ and $20$ (dashed lines).   }
\label{Fig2}
\end{figure}

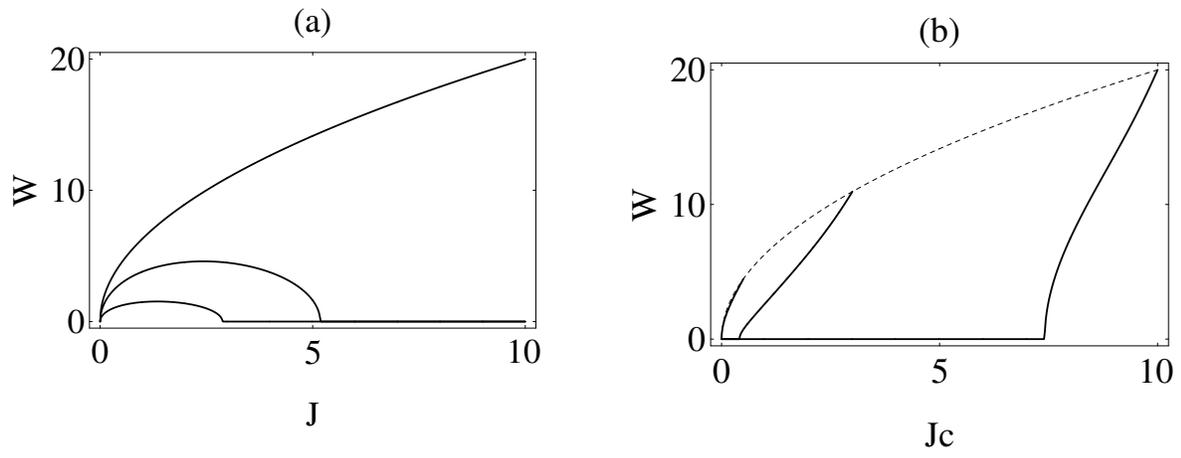
\begin{figure}
\caption{Critical intrachain disorder $W=(12t_x/\tau )^{1/2}$ [$t_x=1$]. 
(a) vs $J$ for $c=1$, $0.5$ and $0.1$
(from top to bottom). (b) vs $c$ for $J=0.5$, $3$ and $10$ (from left
to right). Dashed-line: $W=(24t_x\bar t_\perp /\gamma )^{1/2}$ vs $Jc$. }
\label{Fig3}
\end{figure}

For $M=1$, the position of the MIT strongly depends on the value of
$\gamma $. For $\gamma <1/2$, $\rho $ is a decreasing monotonous function of
$\alpha $. For $\gamma >1/2$, $\rho $  decreases when $\alpha \lesssim 1$ but
increases when $\alpha \gtrsim 1$. Choosing $\gamma >1/2$, we reproduce the
numerical results of ZEE which shows a non-monotonous behavior of $\rho $ as
a function of $\alpha $ (Fig.\ \ref{Fig2}). When $\alpha \ll 1$, $\tau '\gg
\tau $ on the transition line. The main source of scattering comes from the
intrachain disorder. The main effect of an increase of $\alpha $ is then to
increase the effective transverse diffusion coefficient $\tilde D''_\perp
+1/\tau '$, which favors the delocalized phase. When $\alpha
\gtrsim 1$, the random interchain coupling becomes the main source of
scattering ($\tau '\lesssim \tau $). As shown in Fig.\ \ref{Fig2}, for
$\gamma >1/2$ this effect dominates over the increase of the
effective transverse diffusion coefficient and favors the 
localized phase. We can also 
obtain from (\ref{MIT1}) the critical intrachain disorder 
$W=(12t_x/\tau )^{1/2}$ as a function of $J$ ($c$ fixed) or $c$
($J$ fixed) (Fig.\ \ref{Fig3}). (The intrachain transfer integral $t_x$ is
related to the Fermi velocity by $v_F=2t_xa$ for a half-filled band.)   
The very good agreement between our results and the numerical results of ZEE
(compare Figs.\ \ref{Fig2},\ref{Fig3} with Ref.\ \cite{Zambetaki}) 
strongly supports the validity of our approach. 

For $M\gg 1$, the position of the MIT does not
depend on the precise value of $\gamma $ (which should be of order
unity). Contrary to the case $M=1$, the main source of scattering on the
transition line remains the intrawire disorder even when $\alpha >1$. 
We find that $\rho $ is a decreasing monotonous function 
of $\alpha $ (Fig.\ \ref{Fig2}). 
This result is in agreement with PE. However, the position of the transition
does not correspond exactly to the one obtained by PE. In
particular, we find that $\rho $ approaches zero when $\alpha \gg 1$ while
PE obtained $\rho \sim 1$ in this limit. Moreover, since the transition is
described by a NL$\sigma $M, we expect the critical behavior to be the
usual one \cite{Belitz94}, in agreement with the numerical calculation of
ZEE but in disagreement with PE's results. 

I am grateful to G. Montambaux and J.P. Pouget for useful
discussions. I also wish to thank I. Zambetaki for sending me Ref.\
\cite{Zambetaki} prior to publication, C. Bourbonnais for discussions on the
work reported in Ref.\ \cite{Boies95} and D. Boies for sending me a copy of
his PhD thesis.

\begin{references}

\bibitem{poly} For a description of transport properties of highly
conducting doped polymers, see contributions in Synth. Met. {\bf 65} (1994)
and the Proceedings of the International Conference on Science  and
Technology of Synthetic Metals (ICSM'94), Synth. Met. {\bf 69} (1994). 

\bibitem{Q1D} See for instance S. Kivelson and A.J. Heeger, Synth. Met. {\bf
22}, 371 (1988); Z.H. Wang, Phys. Rev. B {\bf 43}, 4373 (1991);
H.H.S. Javadi {\it et al.}, Phys. Rev. B {\bf 43}, 2183 (1991);
V.N. Prigodin and S. Roth, Synth. Met. {\bf 53}, 237 (1993).  

\bibitem{Prigodin93} V.N. Prigodin and K.B. Efetov, Phys. Rev. Lett. {\bf
70}, 2932 (1993); Synth. Met. {\bf 65}, 195 (1994). 

\bibitem{Zambetaki} I. Zambetaki, S.N. Evangelou and E.N. Economou,
preprint.   

\bibitem{Boies95} A related approach has been used recently to study the
instabilities of coupled Luttinger liquids: D. Boies, C. Bourbonnais and
A.-M.S. Tremblay, Phys. Rev. Lett. {\bf 74}, 968 (1995); D. Boies, PhD
thesis, Sherbrooke (unpublished). 

\bibitem{Belitz94} For a review on the NL$\sigma $M approach to Anderson
localization, see D. Belitz and T.R. Kirkpatrick, Rev. Mod. Phys. {\bf 66},
261 (1994).

\bibitem{Wolfle84} P. W\"olfle and R.N. Bhatt, Phys. Rev. B {\bf 30}, 3542
(1984). 

\bibitem{Prigodin83} V.N. Prigodin and Yu.A. Firsov, JETP Lett. {\bf 38},
284 (1983); Yu.A. Firsov, in {\it Localization and Metal-Insulator
Transition}, edited by H. Fritzshe and D. Adler (Plenum, New York, 1985). 

\bibitem{Vollhard80} D. Vollhard and P. W\"olfle, Phys. Rev. B {\bf 22},
4666 (1980). 

\bibitem{Dorokhov83} O.N. Dorokhov, Solid state Comm. {\bf 46}, 605 (1983);
{\bf 51}, 381 (1984); N.A. Panagiotidis, S.N. Evangelou, and G. Theodorou,
Phys. Rev. B {\bf 49}, 14122 (1994). 

\bibitem{Dupuis92} W. Apel and T.M. Rice, J. Phys. C {\bf 16}, L1151 (1983);
N. Dupuis and G. Montambaux, Phys. Rev. B {\bf 46}, 9603 (1992). 

\end{references}
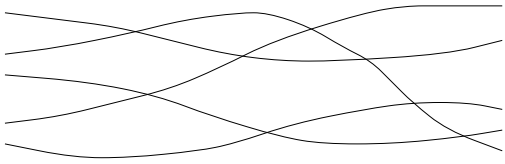
\end{document}